\documentclass[aps,prl,showpacs,amsmath,amssymb,amsfonts,lengthcheck,longbibliography,superscriptaddress]{revtex4-2}

\usepackage{graphicx}\graphicspath{ {figures/} }
\usepackage[caption=false]{subfig}
\usepackage[colorlinks=true,allcolors=blue]{hyperref}
\usepackage{braket}

\newcommand{\pd}{\partial}
\newcommand{\tr}[1]{\text{tr}\left\{#1\right\}}
\newcommand{\avg}[1]{\left\langle #1 \right\rangle}

\begin{document}

\title{Shortcuts to thermodynamic quasistaticity}

\author{Artur Soriani}
\email{asorianialves@gmail.com}

\author{Eduardo Miranda}
\affiliation{Gleb Wataghin Institute of Physics, University of Campinas, Campinas, São Paulo 13083--950, Brazil}

\author{Sebastian Deffner}
\affiliation{Department of Physics, University of Maryland, Baltimore County, Baltimore, Maryland 21250, USA}

\author{Marcus V. S. Bonan\c{c}a}
\affiliation{Gleb Wataghin Institute of Physics, University of Campinas, Campinas, São Paulo 13083--950, Brazil}

\date{\today}

\begin{abstract}
The operation of near-term quantum technologies requires the development of feasible, implementable, and robust strategies of controlling complex many body systems.
To this end, a variety of techniques, so-called ``shortcuts to adiabaticty'', have been developed.
Many of these shortcuts have already been demonstrated to be powerful and implementable in distinct scenarios.
Yet, it is often also desirable to have additional, approximate strategies available, that are applicable to a large class of systems.
In this work, we hence take inspiration from thermodynamics and propose to focus on the macrostate, rather than the microstate.
Adiabatic dynamics can then be identified as such processes that preserve the equation of state, and systematic corrections are obtained from adiabatic perturbation theory.
We demonstrate this approach by improving upon fast quasiadiabatic driving, and by applying the method to the quantum Ising chain in the transverse field.
\end{abstract}

\maketitle

The word ``adiabatic'' is derived from the Greek \emph{adiabatos}, which means literally ``impassable''.
In thermodynamics, an adiabatic constraint is a ``wall'' that is impassable to heat, and thus an adiabatic process is a thermodynamic state transformation during which no heat is exchanged \cite{Callen1985}.
However, the notion of adiabaticity has found a much broader application in Hamiltonian dynamics \cite{Goldstein1980}.
In classical mechanics, an ``adiabatic invariant'' is any quantity that remains constant under the Hamiltonian equations of motion, given infinitely slow variations of the Hamiltonian \cite{Goldstein1980}.

This insight led Born to the formulation of the quantum adiabatic theorem \cite{Born1928}, which states that, during infinitely slow variation of the Hamiltonian, no transitions between energy levels occur.
Obviously, such adiabatic processes are highly desirable in quantum technological applications.
Recent years have seen tremendous research efforts in facilitating such excitation-free processes with finite time driving.
Under the umbrella of \emph{shortcuts to adiabaticity} (STA) \cite{Torrontegui2013,Guery-Odelin2019} a large variety of techniques has been developed, of which counterdiabatic driving \cite{Demirplak2003,Demirplak2005,Berry2009,Deffner2014PRX,Iram2021NP,Ilker2022PRX}, invariant based inverse engineering protocols \cite{Lewis1969,Chen2010,Torrontegui2014,Kiely2015,Levy2018NJP}, and the fast-forward technique \cite{Masuda2010,Masuda2011,Torrontegui2012,Masuda2014,Deffner2015NJP,Jarzynski2017,Myers2021PRQ} have arguably received the most attention, with applications in vastly different physical scenarios. For instance, counterdiabatic driving is particularly well-suited to optimally control the dynamics of cold ion traps \cite{An2016NC,Funo2017PRL}. However, implementing STA in more complex quantum system can become rather involved \cite{Campo2012PRL,Campbell2015PRL,Balasubramanian2018PRA,Cohn2018NJP,Ness2018NJP,Carolan2022PRA}. Thus, it appears very desirable to find alternative and approximate schemes, that may provide more universally applicable control strategies. This has already led to the development of ``resource friendly'' control strategies \cite{Sels2017,Claeys2019,Passarelli2020,Passarelli2022,Hegade2021,Mbeng2022}, that provide alternative means to suppress excitations arising from populating energetically high-lying microstates.

One of the main causes for the complexity of finding realistically useful STA rests in the fact that, to a certain degree, all methods originate in circumventing the quantum adiabatic theorem \cite{Born1928}.
Hence, the focus is on preserving the occupation probabilities of the energy eigenstates, i.e., microstates \cite{Callen1985}.
However, in most experimental settings quantum states cannot be easily measured; rather, \emph{thermodynamic observables} are monitored.  Therefore, \emph{thermodynamic control} has been suggested as a possible way to construct approximate STA \cite{martinezNP2016}, see Ref.~\cite{Deffner2020EPL} for a recent perspective.
However, thermodynamic control methods are usually applied with a focus on lowering the energetic cost of a given thermodynamic process \cite{liPRE2017,chenPRE2019,pancottiPRX2020,liPRL2022,frimPRE2022}.

In the present letter, we change the paradigm of this approach by proposing genuine \emph{shortcuts to thermodynamic quasistaticity}.
To this end, we fully accept the thermodynamic mind set, namely, we seek STA that preserve the \emph{adiabatic macrostate} and not the occupations of microscopic energy eigenstates of a quantum system.
Hence, we demand that the macrostate of a driven system (approximately) fulfills an instantaneous equation of state.
Such a control strategy is constructed by exploiting \emph{adiabatic perturbation theory} \cite{Rigolin2008}, which has recently proven powerful in assessing nonequilibrium excitations in driven quantum Ising chains \cite{Soriani2022PRA_lin,Soriani2022PRA}.
To demonstrate the versatility of the approach, we benchmark our results against other STA, in particular, against \emph{fast quasiadiabatic driving} \cite{Kastberg1995,Bowler2012,Martinez2013,Martinez2015}, which is closest in spirit to our approach.

\paragraph*{Preliminaries}

We start by establishing notions and notations.
Consider a quantum system described by a Hamiltonian $H(\lambda) = \sum_n E_n(\lambda) \ket{n(\lambda)} \bra{n(\lambda)}$, where $E_n(\lambda)$ and $\ket{n(\lambda)}$ are parametric, nondegenerate eigenvalues and eigenstates, respectively.
Moreover, $\lambda$ is an external control parameter, such as the volume of a gas container or a magnetic field.
In the following, we will be interested in thermodynamic state transformations that are driven by varying $\lambda = \lambda(t)$ (also called a protocol), between times $t_i$ and $t_f$, taking the external parameter from an initial value $\lambda_i$ to a final value $\lambda_f$.
Moreover, we assume that the quantum system is thermally insulated, and therefore, its time evolution is unitary.
Note that unitary dynamics are necessarily thermodynamically adiabatic in the traditional sense, since no heat is exchanged.
Thus, unless otherwise stated, ``adiabatic'' means ``quasistatic'' henceforth.

We further assume that the system is initially prepared in a quantum state that is diagonal in the energy eigenbasis, $\rho_i = \sum_n p_n \ket{n_i} \bra{n_i}$, where the subscript $i$ means that a given quantity is evaluated at $t_i$, and $\ket{n_i}=\ket{n(\lambda_i)}$.
The time-dependent state is then determined by the von Neumann equation, $i\hbar\, \dot{\rho}(t) = [ H(\lambda) , \rho(t) ]$, and we denote derivatives with respect to time by a dot.

It is worth emphasizing that, even if the initial state, $\rho_i$, is chosen to be an equilibrium state, $\rho(t>t_i)$ may be arbitrarily far from equilibrium.
Given an initially canonical state [$\rho_i \propto \exp{\left(-\beta H_i\right)}$], even an infinitely slow process will generally not keep the system in canonical equilibrium.
This is because the quasistatic evolution preserves the statistical weights in the initial Hamiltonian.
However, in the present analysis, our main focus is also not the microstate, but rather the thermodynamic macrostate.

In (quantum) thermodynamics, a macrostate is fully characterized by its state variables \cite{Callen1985,Deffner2019book}, which fulfill an equation of state (EOS).
At any instant, the EOS can be obtained by calculating the equilibrium average of the generalized force, $F(\lambda)$, which is given by \cite{Callen1985}
\begin{equation}
 \label{eq:GeneralizedForce}
F(\lambda) = - \frac{\pd H(\lambda)}{\pd \lambda},
\end{equation}
and $\Lambda \equiv \tr{\rho F}$ is the state variable conjugate to $\lambda$.
For any driven process, and writing the time-dependent quantum state as $\rho(t)=\sum_n p_n \ket{\psi_n(t)}\bra{\psi_n(t)}$, the corresponding average generalized force reads
\begin{equation} \label{eq:AverageForce}
\Lambda(t) = \sum_{n} p_n \braket{\psi_n(t) | F(\lambda) | \psi_n(t)}.
\end{equation}
Here, $\ket{\psi_n(t)}$ is a solution of the corresponding Schr\"odinger equation.

\paragraph*{Thermodynamic state transformations}

Before we analyze the more general out of equilibrium situation, we inspect Eq.~\eqref{eq:AverageForce} in the adiabatic limit $\tau \to \infty$.
The adiabatic theorem dictates that, if the evolution is slow enough, the solution to Schrödinger's equation can be written as \cite{Messiah1962book}
\begin{equation}
 \label{eq:AdiabaticLimit}
\ket{\psi_n^{(0)}(t)} = e^{i \phi_n(t)} \ket{n(\lambda)},
\end{equation}
where the superscript $(0)$ denotes the adiabatic limit and $\phi_n(t)$ is the usual adiabatic phase (dynamic plus geometric).
In this case, Eq.~\eqref{eq:AverageForce} simplifies to
\begin{equation} 
\label{eq:EquationOfState}
\Lambda^{(0)} = \sum_n p_n F_{nn}(\lambda),
\end{equation}
where $F_{mn}(\lambda) = \braket{ m(\lambda) | F(\lambda) | n(\lambda) }$.
Notice the lack of explicit time dependence in Eq.~\eqref{eq:EquationOfState}: this is the conventional EOS.
For infinitely slow variations of $\lambda$, Eq.~\eqref{eq:EquationOfState} describes the evolution of the macroscopic state in any mechanically adiabatic (and thermodynamically adiabatic) process, i.e., for a thermodynamic state transformation.

\paragraph{Beyond the adiabatic limit}

Using adiabatic perturbation theory (APT), whose details we leave for the Supplemental material \cite{SM}, we can systematically compute finite-time corrections to the EOS \eqref{eq:EquationOfState}.
Using Eqs.~(1)--(3) of the Supplemental Material \cite{SM} in Eq.~\eqref{eq:AverageForce} and keeping terms up to $\mathcal{O}(\tau^{-1})$, the first-order correction becomes
\begin{equation}
\begin{split}
\label{eq:AverageForceFirstOrder}
\Lambda^{(1)}(t) & = \underset{m \neq n}{\sum_{m,n}} p_n \Re \left\{2 C_{mn}^{(1)}(t) F_{mn}^*(\lambda) \right\} \\
& = 2\hbar \dot{\lambda}_i \underset{m \neq n}{\sum_{m,n}} p_n \Im\left\{ F_{mn,i} \frac{e^{i \phi_{mn}(t)}}{E_{mn,i}^2} F_{mn}^*(\lambda)  \right\},
\end{split}
\end{equation}
where we used the fact that the product of $F_{mn}^*(\lambda)$ and the first term of Eq.~(2) of the Supplemental Material \cite{SM} is purely imaginary.
We immediately observe that the first-order correction to the EOS is directly proportional to the time derivative of the external parameter \emph{at the beginning of the process}.
Hence, for all protocols with $\dot{\lambda}_i = 0$, the EOS is preserved up to $\mathcal{O}(\tau^{-2})$ in any sufficiently slow process.
We stress that this conclusion is independent of the Hamiltonian considered, only depending on the validity of APT. Thus, we have unveiled a universal design principle for optimal control strategies applicable in any gapped quantum system, simple as well as complex.

Strategies where the time derivatives of the protocols vanish at the end points of the evolution have already been discussed as ways to guarantee adiabaticity in the microstate~\cite{Jansen2007,Morita2008,Rezakhani2010,Venuti2018}.
However, we emphasize that the first-order result for the macrostate only depends on the initial derivative, and not the final derivative.
This still leaves a lot of freedom in finding ``optimal'' and experimentally implementable protocols.
Thus, it should be obvious that even better results can be achieved by complementing our macroscopic strategy with microscopic methods.

\begin{figure*}
\subfloat[\label{fig:TI_para_EOS}]{\includegraphics[width=.5\textwidth]{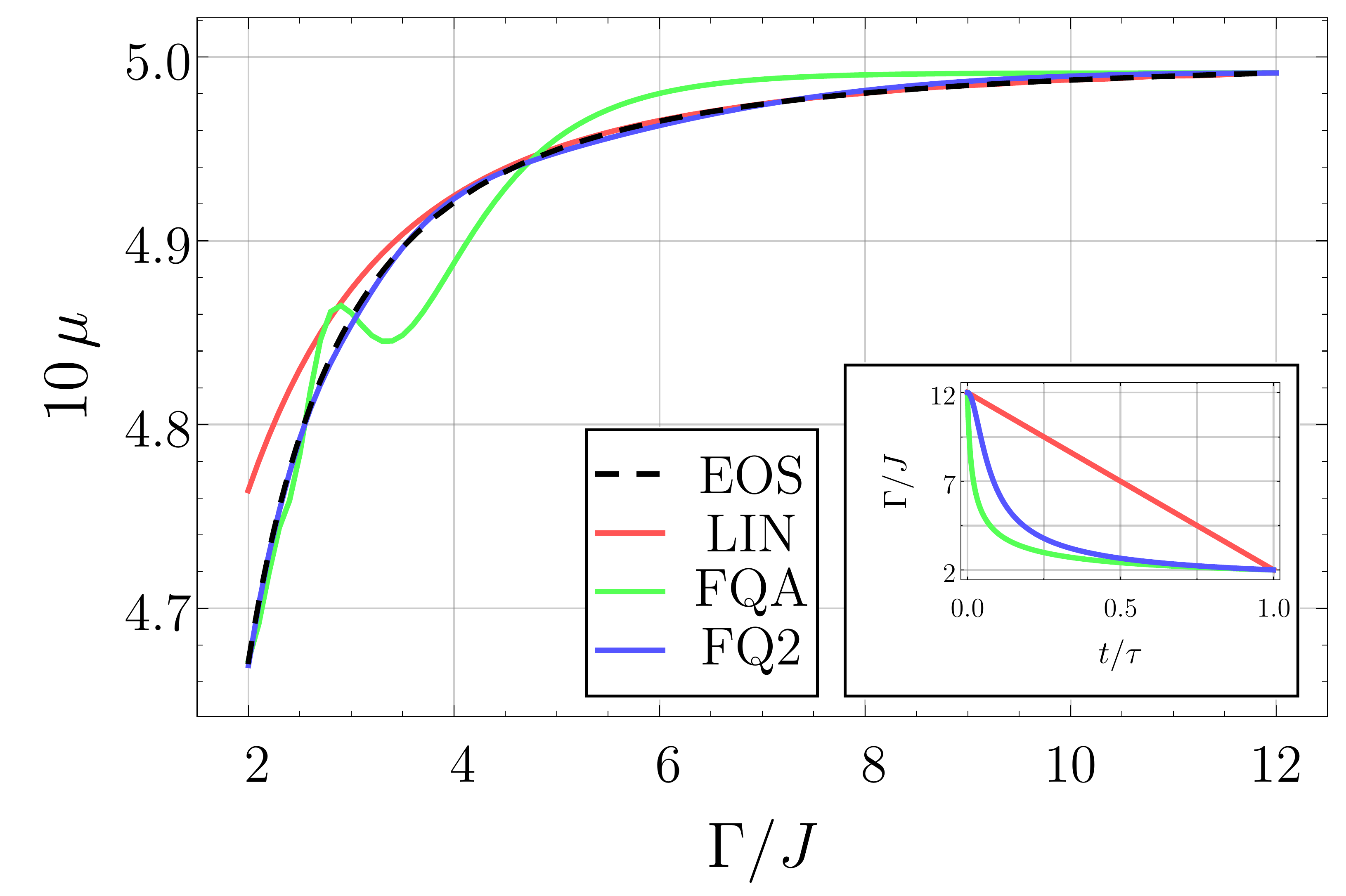}}
\subfloat[\label{fig:TI_para_force}]{\includegraphics[width=.5\textwidth]{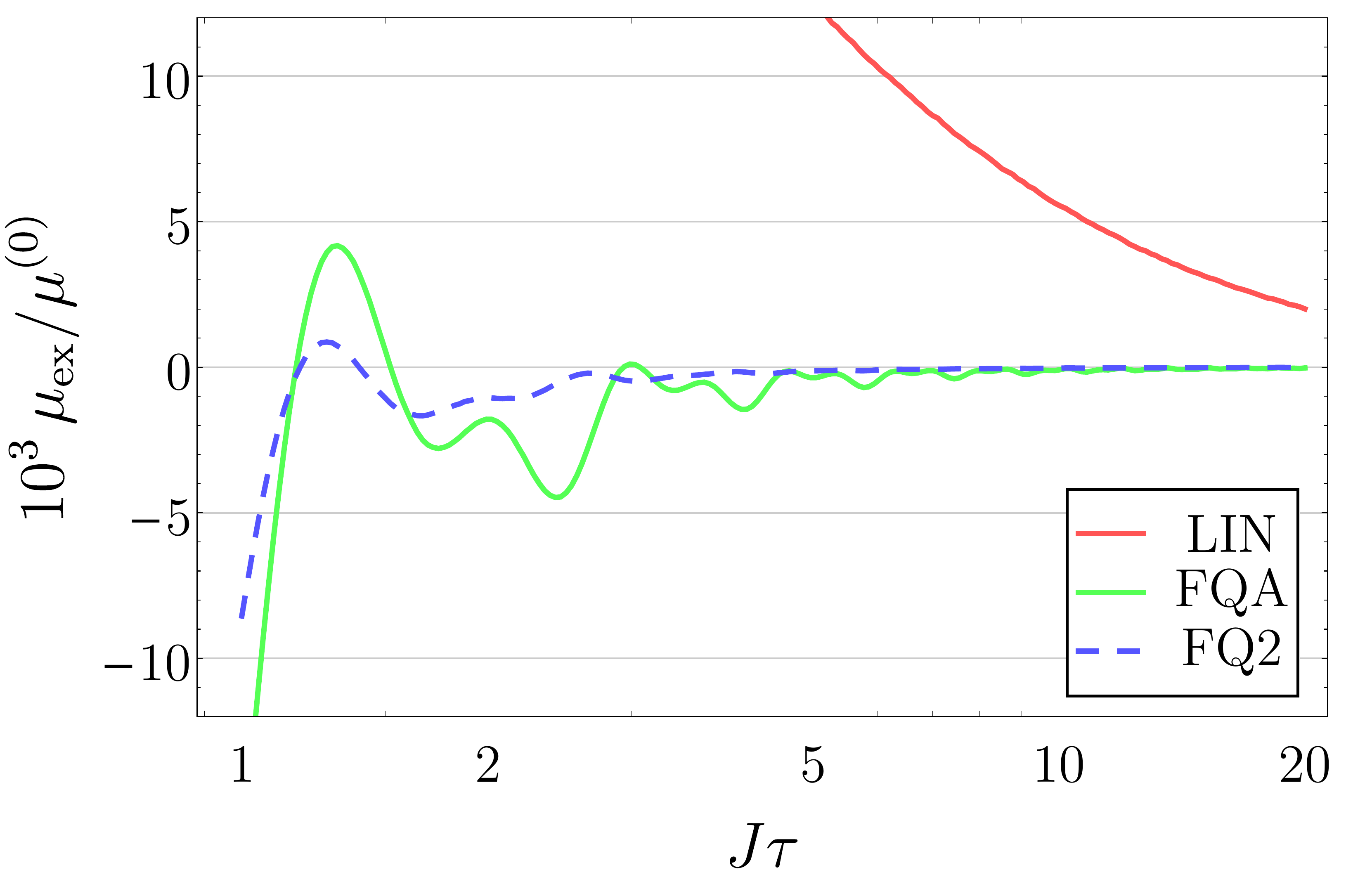}}
\caption{\label{fig:TI_para}
Magnetization of the TI chain in the entirely paramagnetic process with $N = 100$ starting at zero temperature.
The results were numerically obtained from the exact time-dependent dynamics.
(a) State diagram of the TI chain for an adiabatic (quasistatic) evolution (EOS), the LIN, the FQA, and the FQ2 protocols for $J \tau = 3$, starting from the top right corner.
The inset shows the time dependence of each protocol.
(b) Excess magnetization $\mu_\mathrm{ex} = \mu - \mu^{(0)}$ at the end of the process vs process duration.
}
\end{figure*}

\paragraph*{Fast quasiadiabatic driving}

One strategy to ensure APT convergence is the application of fast quasiadiabatic (FQA) protocols \cite{Kastberg1995,Bowler2012,Martinez2013,Martinez2015} and related approaches \cite{Guery-Odelin2019}.
If there is only one relevant energy gap $E_{mn}(\lambda)$ in the quantum system, FQA provides a protocol $\lambda(t)$ for which first-order APT transitions between eigenstates $m$ and $n$ are equally likely at any instant.
This protocol is the solution to a first order differential equation \cite{Kastberg1995,Bowler2012,Martinez2013,Martinez2015}
\begin{equation} 
\label{eq:FQADifferentialEquation}
\hbar \left| \frac{ \dot{\lambda}(t) F_{mn}(\lambda) }{E_{mn}^2(\lambda)} \right| = c_1,
\end{equation}
where $c_1$ is a constant that, together with the integration constant, is uniquely defined by the boundary conditions $\lambda(t_i) = \lambda_i$ and $\lambda(t_f) = \lambda_f$.
For a generic protocol, microscopic adiabaticity is secured if the left-hand side of Eq.~\eqref{eq:FQADifferentialEquation} is much smaller than unity for any $t$, the quantitative adiabatic condition \cite{Tong2005,Tong2010} [Eq.~(4) of the Supplemental Material \cite{SM}].
The boundary conditions always lead to $c_1 \propto \tau^{-1}$, which means that the FQA protocol still requires large enough $\tau$ for the adiabatic condition to be fulfilled.
FQA's advantage is that it \emph{naturally slows down} where $E_{mn}(\lambda)$ is small [see Eq.~\eqref{eq:FQADifferentialEquation}], and thus, it may reach the adiabatic condition and make APT converge for a smaller $\tau$, when compared to a generic protocol.

Curiously, FQA is limited to suppressing first-order transitions.
The authors of Ref.~\cite{Martinez2015} remark that considering transitions of higher-than-one order APT is not possible, since the associated differential equation would not have enough constants to satisfy the boundary conditions on $\lambda$ and its derivatives.
For example, demanding the second-order APT transition probabilities to be uniform along the process gives a second-order differential equation,
\begin{equation} 
\label{eq:FQ2DifferentialEquation}
\hbar^2 \left| \frac{1}{E_{mn}(\lambda)} \frac{d}{dt} \left( \frac{ \dot{\lambda}(t) F_{mn}(\lambda) }{E_{mn}^2(\lambda)} \right) \right| = c_2,
\end{equation}
which was obtained from Eq.~\eqref{eq:FQADifferentialEquation} with the proper substitution to second order coefficients, discussed in the Supplemental Material \cite{SM}.
The three available constants ($c_2$ plus two integration constants) in the solution of Eq.~\eqref{eq:FQ2DifferentialEquation} are insufficient to satisfy the four boundary conditions --- two on $\lambda$ (same as FQA) and two on $\dot{\lambda}$, which are necessary to make the second-order APT correction be the relevant correction.

Above, we have seen that, from the macroscopic dynamics, Eq.~\eqref{eq:AverageForceFirstOrder}, optimal driving protocols obey $\dot{\lambda}=0$ at the beginning (and not at the end).
This additional condition permits us to uniquely solve Eq.~\eqref{eq:FQ2DifferentialEquation}, if we impose the same boundary conditions as the FQA method \emph{plus} $\dot{\lambda}(t_i) = 0$, which leads to $c_2 \propto \tau^{-2}$.
We will be referring to this strategy as FQ2, and as we will see shortly, FQ2 clearly outperforms FQA.
We once again bring attention to the fact that making $\dot{\lambda}(t_i) = 0$ gives null first order APT correction for the EOS of \emph{any gapped system}.
Equations \eqref{eq:FQADifferentialEquation} and \eqref{eq:FQ2DifferentialEquation}, which do depend on the system through its eigenspectrum, are primarily used to guarantee early APT validity and can be applied even when the Hamiltonian is only numerically diagonalizable.
In fact, at low temperature, knowledge of only a few eigenlevels may be necessary, since only transitions between the lowest energy eigenstates are relevant (see Fig.~2 of the Supplemental Material \cite{SM}).

\paragraph*{Illustrative example: quantum Ising chain}

We now apply the above developed strategy to control a thermodynamically relevant, exactly solvable system: the transverse field Ising model (TI) \cite{Pfeuty1970,Dziarmaga2005}.
The Hamiltonian reads
\begin{equation} 
\label{eq:TIHamiltonian}
H_{\mathrm{TI}}(\Gamma) = - \frac{1}{2} \left( J \sum_{j=1}^{N} \sigma^z_j \sigma^z_{j+1} + \Gamma \sum_{j=1}^N \sigma^x_j \right),
\end{equation}
where $J$ is the coupling constant, $\Gamma$ is the external magnetic field and $\sigma_j^{x,z}$ are standard Pauli matrices for each spin $j$ (with periodic boundary conditions).
In the thermodynamic limit $N \to \infty$, this system displays a quantum critical point (QCP) at $\Gamma = J$, where the energy gap between ground and first excited states vanishes.
For simplicity, we assume $N$ to be even and that the system is initially prepared in its ground state.
The force is $F_{\mathrm{TI}} = \sum_{j=1}^N \sigma^x_j/2$, while the nonequilibrium magnetization per spin reads
\begin{equation} 
\label{eq:TIMagnetization}
\mu(t) = \frac{1}{2N} \sum_{j=1}^N \avg{\sigma^x_j}(t).
\end{equation}
In any finite time process, the magnetization can be separated into an adiabatic contribution $\mu^{(0)}$ and an excess contribution $\mu_{\mathrm{ex}}$.
Details for how to calculate the nonequilibrium average in Eq.~\eqref{eq:TIMagnetization} can be found in the Supplemental Material \cite{SM}.

\begin{figure*}
\subfloat[\label{fig:TI_crit_EOS}]{\includegraphics[width=.5\textwidth]{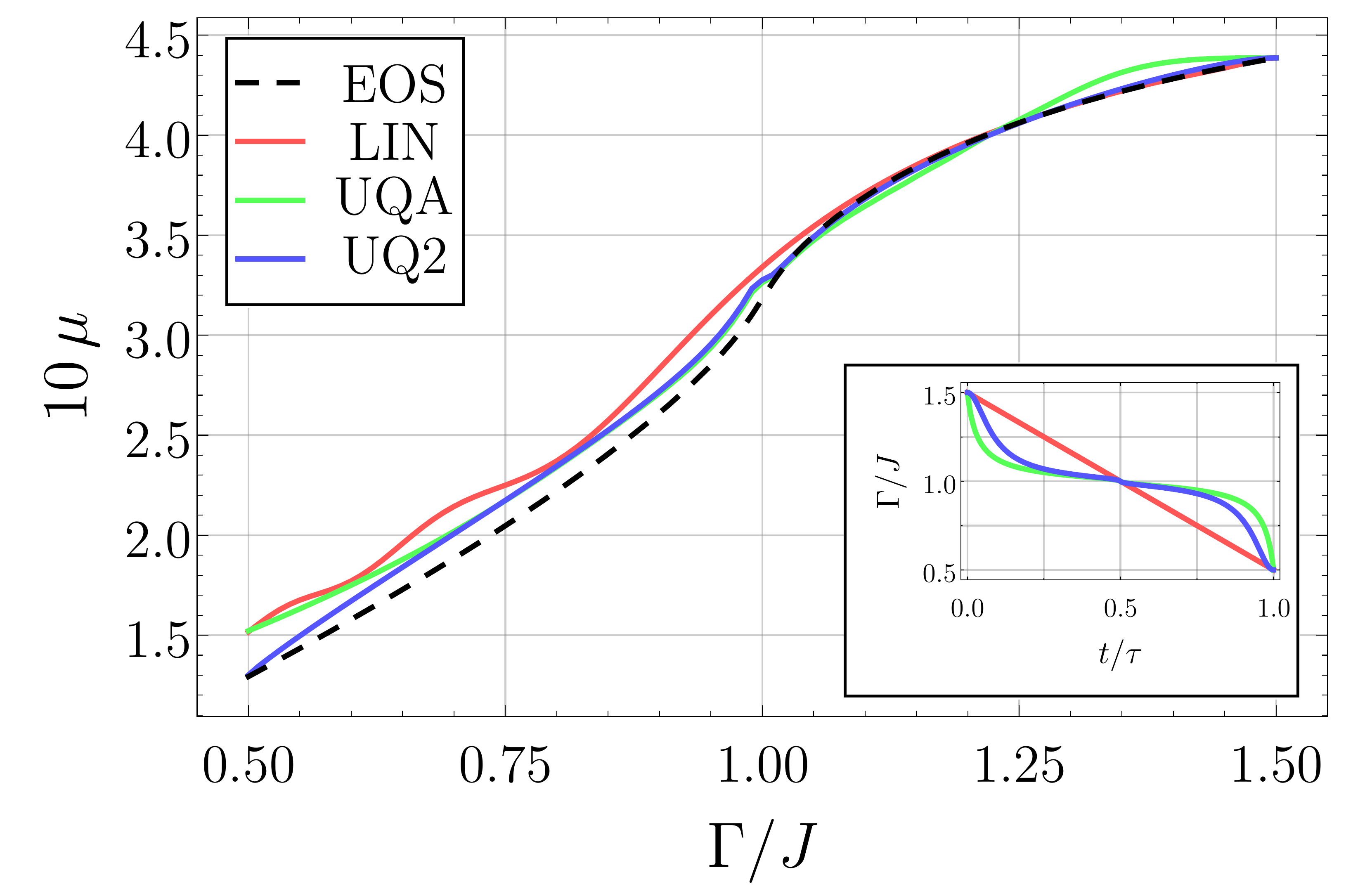}}
\subfloat[\label{fig:TI_crit_force}]{\includegraphics[width=.5\textwidth]{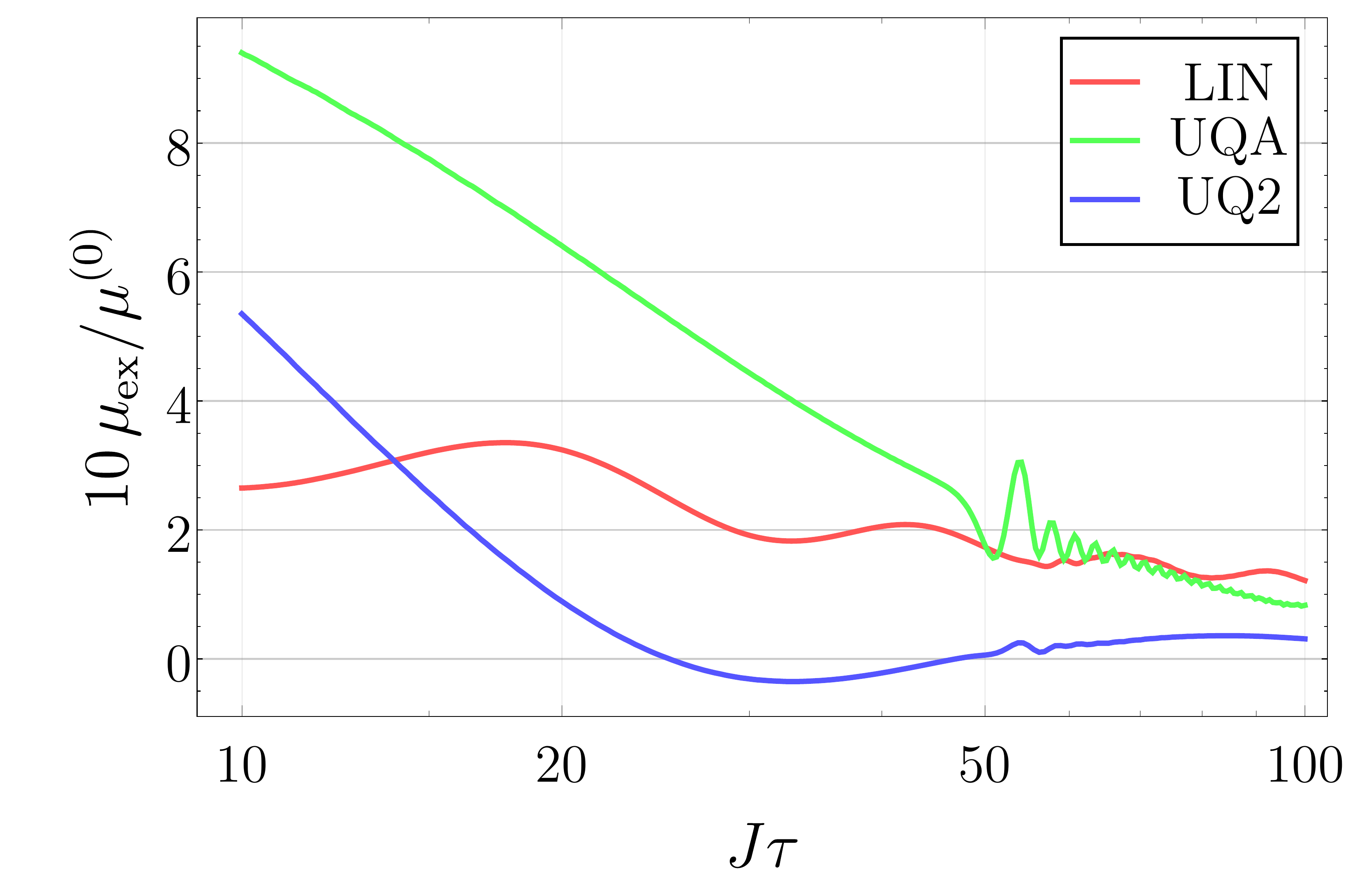}}
\caption{\label{fig:TI_crit}
Magnetization of the TI chain in the QCP crossing process starting at zero temperature with $N = 100$. The results of both panels were numerically obtained from the exact time-dependent dynamics.
(a) State diagram of the TI chain for an adiabatic (quasistatic) evolution (EOS), the LIN, the UQA, and the UQ2 protocols for $J \tau = 50$, starting from the top right corner.
The inset shows the time dependence of each protocol.
(b) Excess magnetization $\mu_\mathrm{ex} = \mu - \mu^{(0)}$ at the end of the process vs process duration.
}
\end{figure*}

First, we consider a process keeping the chain entirely in its paramagnetic phase ($\Gamma > J$) and starting at zero temperature, i.e., with the chain initially prepared in its ground state.
We solve FQA and FQ2 for the smallest gap of the system and compare them to a naive linear protocol (LIN) --- the results for a chain of finite size are shown in Fig.~\ref{fig:TI_para}.
In Fig.~\ref{fig:TI_para_EOS}, we show $\mu$ of Eq.~\eqref{eq:TIMagnetization} vs $\Gamma$ in a process that approaches, but does not cross, the QCP.
The inset contains the time-dependence of each protocol, where it can be seen that both FQA and FQ2 adapt to the system's spectrum, but FQ2 does so while still keeping null first derivative at the start.
FQA has a very high first derivative at the initial time, and this ultimately makes its evolution have notable oscillations around the EOS.
On the other hand, LIN follows the EOS closely, up until a point where the gap gets too small, and it ends up breaking adiabaticity.
Finally, FQ2 follows the EOS right until the end, which is a consequence of its compromise to attain adiabaticity while zeroing the first order correction to the EOS.
In Fig.~\ref{fig:TI_para_force}, we depict the excess magnetization $\mu_{\mathrm{ex}}$ at $t_f$ as a function of $\tau$.
It is clear that FQ2 outperforms FQA for a generic $\tau$, even if FQ2 first crosses the ``adiabatic'' $\mu_{\mathrm{ex}} = 0$ line for a marginally bigger $\tau$ than FQA.

As a second case, we consider the crossing of the QCP, from the paramagnetic phase to the ferromagnetic phase.
In a finite size chain, the gap at the QCP is small but nonzero, which makes adiabaticity difficult but possible to achieve.
In this scenario, the smallness of the energy gap forces the FQA protocol to slow down dramatically around the QCP and, consequently, to speed up around the end points.
This speed-up is detrimental in the ferromagnetic phase of the TI chain, where the gap of many other sub-levels are comparable to the gap of the lowest sub-level (see Fig.~1 of the Supplemental Material \cite{SM}).
Other energy differences can be taken into account when building FQA protocols (see Ref.~\cite{Soriani2022work}), but the associated differential equation is not exactly solvable and hardly numerically solvable when traversing the QCP.
Thus, to circumvent this issue, we apply a similar strategy known as uniform quasi-adiabatic (UQA) \cite{Quan2010} to the lowest sub-level of the TI chain.
It is the solution to Eq.~\eqref{eq:FQADifferentialEquation} with the substitution $F_{mn}(\lambda) \to \pd_{\lambda} E_{mn}(\lambda)$ \cite{Guery-Odelin2019}, motivated by the Kibble-Zurek mechanism of second-order quantum phase transitions.
Thus, we define a UQ2 protocol as the solution of Eq.~\eqref{eq:FQ2DifferentialEquation} with the aforementioned substitution, and we compare it to LIN and UQA in Fig.~\ref{fig:TI_crit}.
Figure \ref{fig:TI_crit_EOS} is the equivalent of Fig.~\ref{fig:TI_para_EOS}, but with a considerably larger process duration, which evidences the difficulty of crossing the QCP while maintaining adiabaticity (in the mechanical sense).
The inset once again shows the time-dependence of each strategy, and it is clear that both UQA and UQ2 slow down around the QCP.
The conclusion is the same as in the paramagnetic process: UQ2 follows the EOS more closely.
Furthermore, as can be seen in Fig.~\ref{fig:TI_crit_force}, UQ2 gives final $\mu_{\mathrm{ex}} = 0$ for a significantly smaller $\tau$ than the other two protocols, which is a consequence of its final first derivative also being null at the end point (see inset of Fig.~\ref{fig:TI_crit_EOS}).

\paragraph*{Concluding remarks}

Controlling complex many body quantum systems is an involved task.
While some strategies have been successfully employed in platforms with great technological promise, such as counterdiabatic driving in ion traps \cite{An2016NC,Funo2017PRL}, more universally applicable paradigms appear desirable.
To this end, we have proposed to take inspiration from the mother of all control theories --- thermodynamics.
Rather than aiming to control the microstate, we have suggested controlling the macrostate and identifying protocols that preserve the equation of state.
This approach is somewhat akin to invariant based strategies \cite{Chen2010,Guery-Odelin2019}, on which we comment in the Supplemental Material \cite{SM}, where we study thermodynamic shortcuts for the driven harmonic oscillator \cite{Husimi1953,Deffner2008,Deffner2010}.
However, our approach significantly goes beyond existing methods, since using adiabatic perturbation theory, finite-time corrections can be systematically computed, which gives systematic conditions for the optimal driving protocols.
The utility of the approach has been demonstrated by improving upon fast quasiadiabatic driving, and its applicability has been demonstrated for the driven Ising chain.

The analyses of state diagrams demonstrate the difference between microscopic adiabaticity and macroscopic adiabaticity.
More specifically, strategies that are better suited for parametric following of microstates (eigenstates) are not necessarily better for parametric following of macrostates (state variables).
It is also worth noting that a notion of relaxation time seems to be absent, which is perhaps expected in isolated systems where relaxation to some sort of equilibrium is not guaranteed.
Nonetheless, there is still the notion of a timescale with which the driving rate must be compared, related to the energy gap between eigenstates.
Last, it is interesting to see that, even though it is possible to stay close to the equation of state in finite time driving, such possibility a does not lead to thermodynamic reversibility.
In other words, applying the same ``optimal'' protocol in the reverse process does not give the same curve in the state diagram as in the forward process and, in fact, the FQ2 strategy we devised to better follow the equation of state does not provide protocols with time-reversal symmetry.

Finally, we note that the present paper fills the gap in a hierarchy of strategies developed for securing adiabaticity in finite time.
First, there are standard shortcuts to adiabaticity, where one seeks to follow the parametric eigenstates of the system.
Second, we have the thermodynamic shortcuts introduced in the present letter, which follow the equation of state.
Third, we have the methods from thermodynamic control, where the focus is on making sure that the energetic cost of a certain manipulation of the system is as close as possible to the cost in a quasistatic process.
It is expected that the further down you go in the hierarchy, the less information is needed to determine the associated optimal driving protocol.

\acknowledgements{
A. S. and M. V. S. B. thank the National Council for Scientific and Technological Development --- CNPq under grant 140549/2018-8 and FAEPEX (Fundo de Apoio ao Ensino, \`a Pesquisa e \`a Extens\~ao)(Brazil)(Grant 2146-22).
M. V. S. B.  also acknowledges the financial support of FAPESP (Funda\c{c}\~ao de Amparo \`a Pesquisa do Estado de S\~ao Paulo) (Brazil)(Grant 2020/02170-4).
E.M. also thanks the support of CNPq through Grant No. 309584/2021-3 and Capes through Grant No. 0899/2018.
S.D. acknowledges support from the U.S. National Science Foundation under Grant No. DMR-2010127.
}

\bibliography{bibliography}

\end{document}


\title{Supplemental Material for ``Shortcuts to thermodynamic quasistaticity''}

\author{Artur Soriani}
\email{asorianialves@gmail.com}

\author{Eduardo Miranda}

\affiliation{Gleb Wataghin Institute of Physics, University of Campinas, Campinas, São Paulo 13083--950, Brazil}

\author{Sebastian Deffner}

\affiliation{Department of Physics, University of Maryland, Baltimore County, Baltimore, Maryland 21250, USA}

\author{Marcus V. S. Bonan\c{c}a}
\affiliation{Gleb Wataghin Institute of Physics, University of Campinas, Campinas, São Paulo 13083--950, Brazil}

\date{\today}

\maketitle

In these Supplementary Materials we provide
(i) an overview of adiabatic perturbation theory,
(ii) further technical details for the driven Ising chain, and
(iii) a comparison of our new method with invariant based reverse engineering for the parametric harmonic oscillator.

\paragraph*{Adiabatic perturbation theory}

(APT) \cite{Rigolin2008} is a perturbative generalization of the adiabatic theorem.
An exact solution of the Schr\"odinger equation can be written as
\begin{equation}
\label{eq:APTDynamicState}
\ket{\psi_n(t)} = e^{i\phi_n(t)} \sum_{p=0}^{\infty} \sum_m C_{mn}^{(p)}(t) \ket{m(\lambda)},
\end{equation}
where $\phi_n(t)$ is the adiabatic phase and the coefficients $C_{mn}^{(p)}(t)$ represent the $p$th order corrections to the adiabatic solution, which is Eq.~(3) of the main text.
For $p=0$, we simply have $C_{mn}^{(0)}(t) = \delta_{mn}$, and Eq.~\eqref{eq:APTDynamicState} truncated at this order reproduces the adiabatic approximation.

For $p>0$, the coefficients $C_{mn}^{(p)}(t)$ can be systematically calculated \cite{Rigolin2008}.
For instance, for $p=1$ and $m \neq n$ we have
\begin{equation} \label{eq:APTFirstOrderTransition}
C_{mn}^{(1)}(t) = i \hbar \left( \frac{M_{mn}(\lambda)}{E_{mn}(\lambda)} - e^{i \phi_{mn}(t)} \frac{M_{mn,i}}{E_{mn,i}} \right),
\end{equation}
where
\begin{equation} \label{eq:MatrixM}
M_{mn}(t) = \dot\lambda(t) \frac{F_{mn}(\lambda)}{E_{mn}(\lambda)},
\end{equation}
$F_{mn}$ are the matrix elements of the generalized force [Eq.~(1) of the main text] and we introduced $E_{mn}(\lambda) \equiv E_m(\lambda) - E_n(\lambda)$ and $\phi_{mn}(t) \equiv \phi_m(t) - \phi_n(t)$.

For the purposes discussed here, higher order coefficients can be obtained from the first order coefficient with iterative substitutions.
For example, $C_{mn}^{(2)}$ with $m \neq n$ is given by Eq.~\eqref{eq:APTFirstOrderTransition} with the change $M_{mn}(t) \to i\hbar \frac{d}{dt} \frac{M_{mn}(t)}{E_{mn}(\lambda)}$.
We can gauge how accurate the theory is by using the so-called quantitative adiabatic condition \cite{Tong2005,Tong2010}
\begin{equation}
\hbar \frac{M_{mn}(t)}{E_{mn}(\lambda)} \ll 1,
\end{equation}
which can be evaluated at any point in time $t$ of a given process.

\paragraph*{Transverse field Ising chain}

\begin{figure}

\includegraphics[width=\columnwidth]{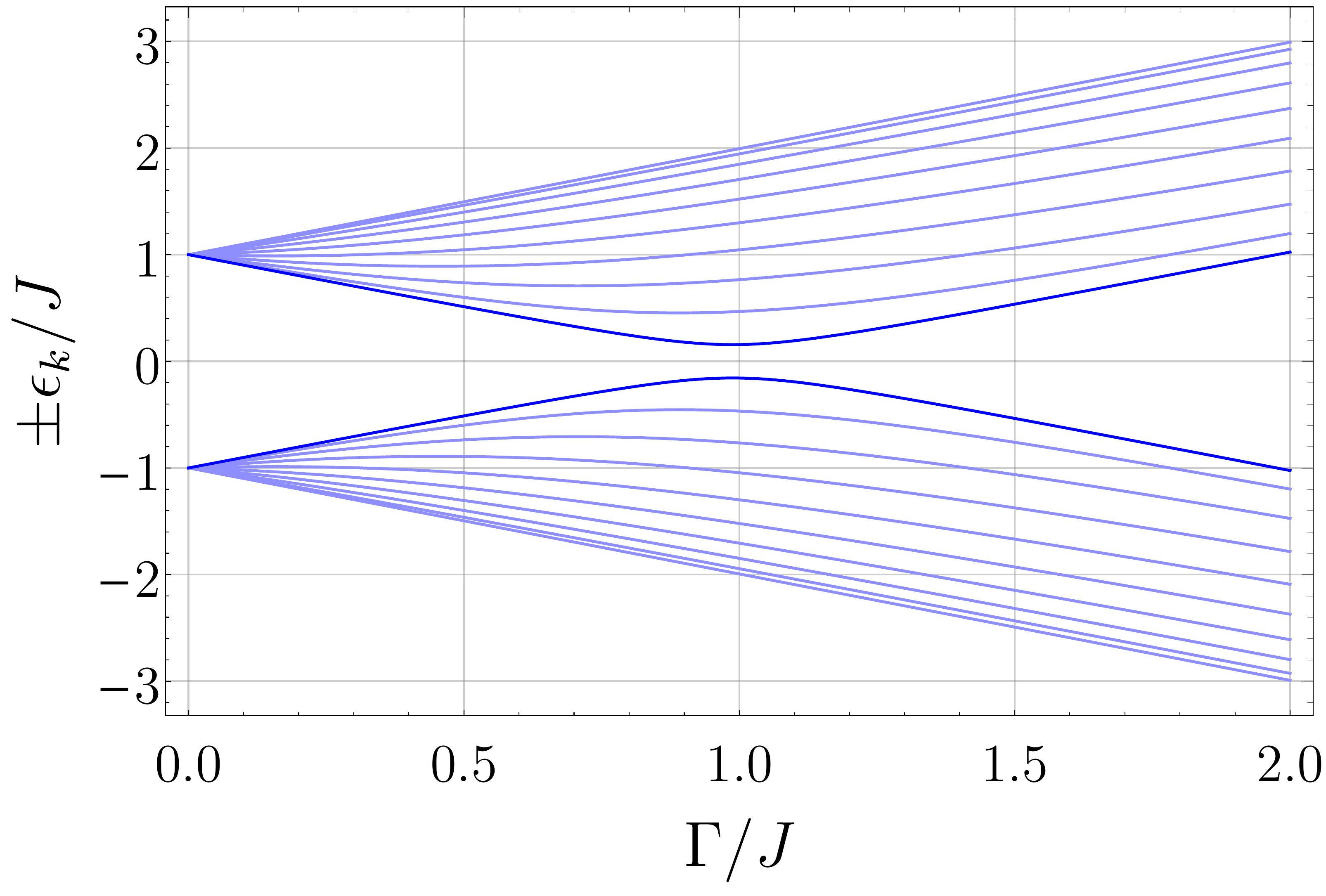}

\caption{\label{fig:TI_EigenEnergies}
Eigen-energies of the transverse field Ising chain [Eq.~\eqref{eq:TIDispersion}] versus the external field $\Gamma$, for $N = 20$ and all possible values of $k$.
}

\end{figure}

The Hamiltonian of the system is given in Eq.~(8) of the main text.
We assume an even number of spins and periodic boundary conditions, while taking $\hbar = 1$.
After a Jordan-Wigner transform, a Fourier transform and a Boguliobov transform \cite{Pfeuty1970}, this Hamiltonian is brought to diagonal form, represented by non-interacting fermions with dispersion
\begin{equation} 
\label{eq:TIDispersion}
\epsilon_k(\Gamma) = \sqrt{\left( \Gamma - J\cos k \right)^2 + J^2 \sin^2 k},
\end{equation}
for $N$ allowed values of momentum
\begin{equation} \label{eq:TIMomentum}
 k_n = (2n + 1) \frac{\pi}{N},
\end{equation}
given integer $n$ between $-N/2$ and $N/2 - 1$.
Figure~\ref{fig:TI_EigenEnergies} shows the energies of Eq.~\eqref{eq:TIDispersion} as a function of $\Gamma$ for $N = 20$, where the energy gap between the two lowest energy levels is seen to shrink at $\Gamma = J$, the quantum critical point (QCP) of the system.

If the system starts the process in its initial ground state, its dynamics can be simplified into the dynamics of $N/2$ two-level systems (known as Landau-Zener systems), one for each positive value of $k$ \cite{Dziarmaga2005}.
The evolved ground state can be written as
\begin{equation} 
\label{eq:TIEvolvedGroundState}
\ket{\psi(t)} = \bigotimes_{k>0} \Big( u_k(t) \ket{\downarrow_k} - v_k(t) \ket{\uparrow_k} \Big),
\end{equation}
where $\ket{\uparrow_k}$ and $\ket{\downarrow_k}$ form a basis of the two-level system labeled by $k$.
Placing Eq.~\eqref{eq:TIEvolvedGroundState} into Schrödinger's equation leads to (omitting the time-dependences of $u_k$, $v_k$ and $\Gamma$)
\begin{equation} 
\label{eq:TIDifferentialEquations}
\begin{split}
i\, \dot{u}_k & =  - \left( \Gamma - J \cos k \right) u_k - J \sin k \, v_k, \\
i\, \dot{v}_k & =  - J \sin k \, u_k + \left( \Gamma - J \cos k \right) v_k.
\end{split}
\end{equation}
The numerical results of presented here and in the main text were obtained from the standard fourth-order Runge--Kutta method applied to Eq.~\eqref{eq:TIDifferentialEquations}.
The magnetization per spin can be expressed as
\begin{equation}
 \label{eq:TIMagnetization}
\mu(t) = \frac{1}{N} \sum_{k>0} \left( \left|u_k\right|^2 - \left|v_k\right|^2 \right).
\end{equation}
where $u_k$ and $v_k$ are calculated from Eq.~\eqref{eq:TIDifferentialEquations}.

\begin{figure}

\includegraphics[width=\columnwidth]{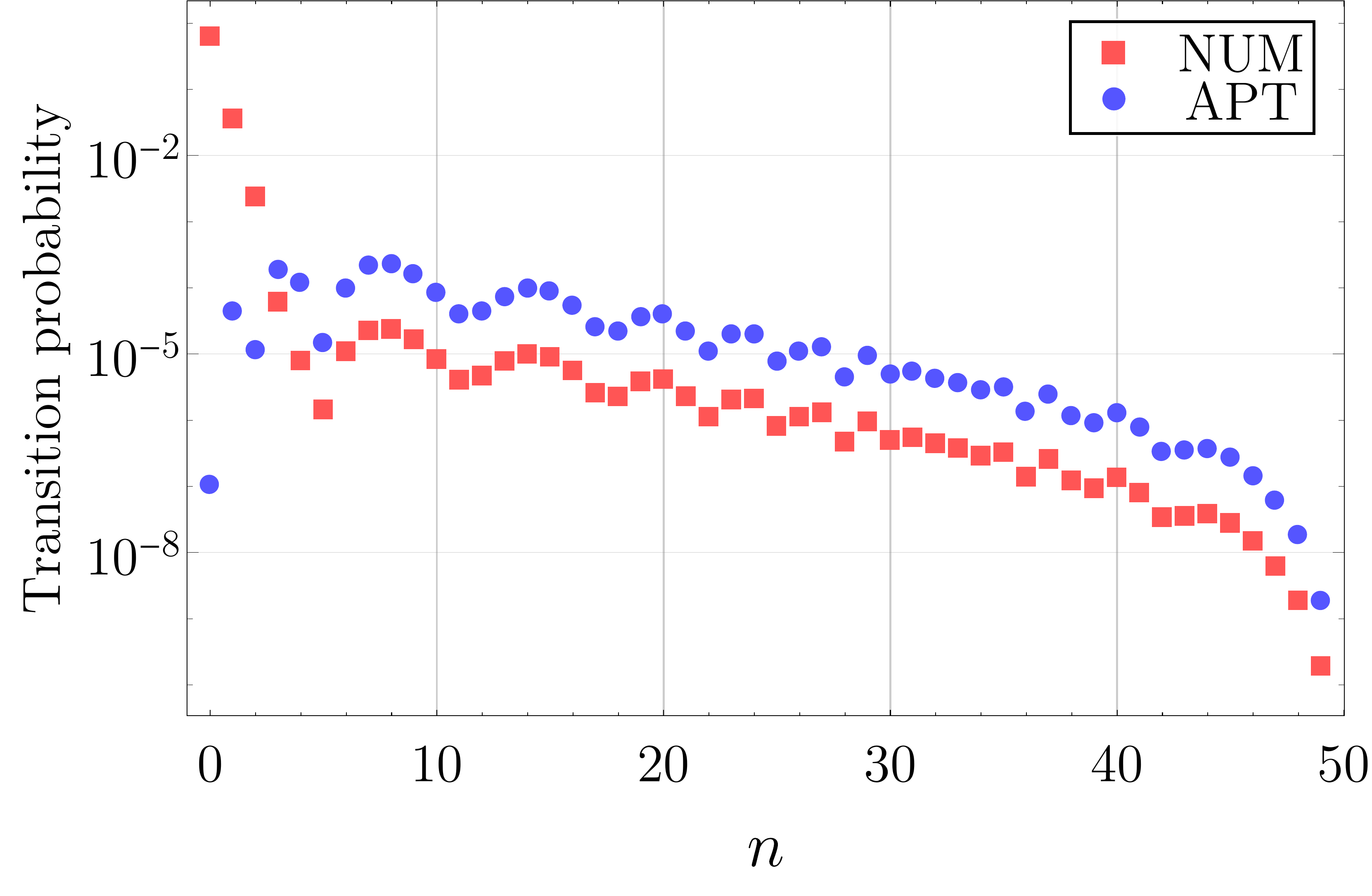}

\caption{\label{fig:TI_transitionProbabilities}
Transition probabilities between the ground state and an excited state with a single pair of fermions with momenta $k_n$ [Eq.~\eqref{eq:TIMomentum}], as a function of $n$, in a process that crosses the QCP of the TI chain with $N = 100$ and $J \tau = 50$.
The red squares represent the result from numerical integration of Eq.~\eqref{eq:TIDifferentialEquations}, while the blue circles represent the analytical result from first-order APT.
}

\end{figure}

Figure \ref{fig:TI_transitionProbabilities} shows the transition probabilities in a process starting in the ground state of the paramagnetic phase and ending in the ferromagnetic phase of the TI chain, the same QCP-crossing process considered in the main text.
In it, we display only transitions to states reached by the creation of a single pair of fermions with momenta $k_n$ [given in Eq.~\eqref{eq:TIMomentum}], which means that the relevant energy gap increases as $n$ (the variable in the horizontal axis) increases.
There is a clear downward trend, which confirms the notion that transition probabilities are higher for pairs of neighbouring energy levels.
For small $n$, the probabilities decrease exponentially with $n$, which means that only the lowest energy eigenstates contribute meaningfully to the dynamics, while the rest of the eigenstates can be ignored in the determination of protocols such as FQA.

\begin{figure*}

\subfloat[\label{fig:TI_backward_para_EOS}]{\includegraphics[width=.5\textwidth]{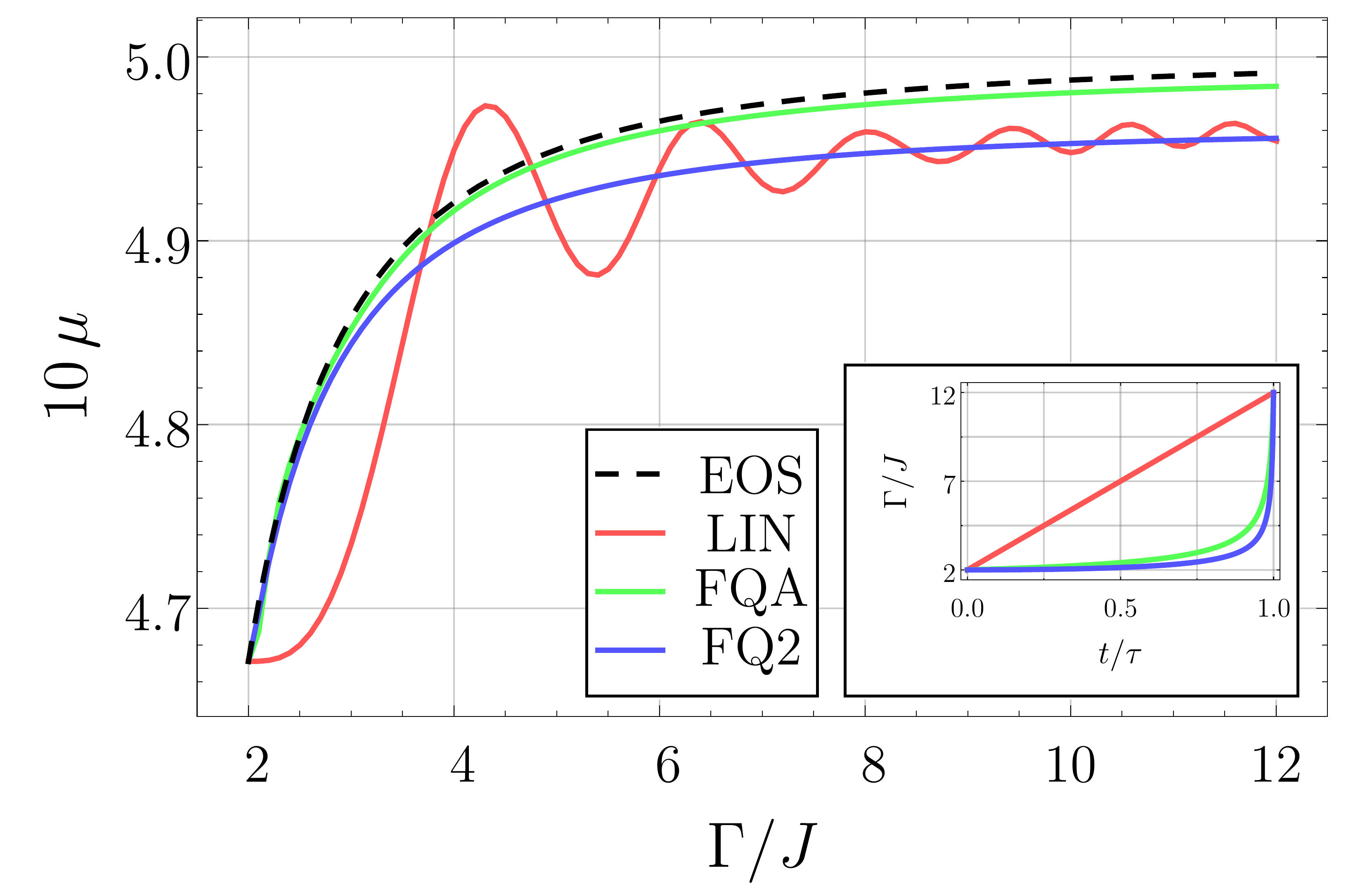}}
\subfloat[\label{fig:TI_backward_crit_EOS}]{\includegraphics[width=.5\textwidth]{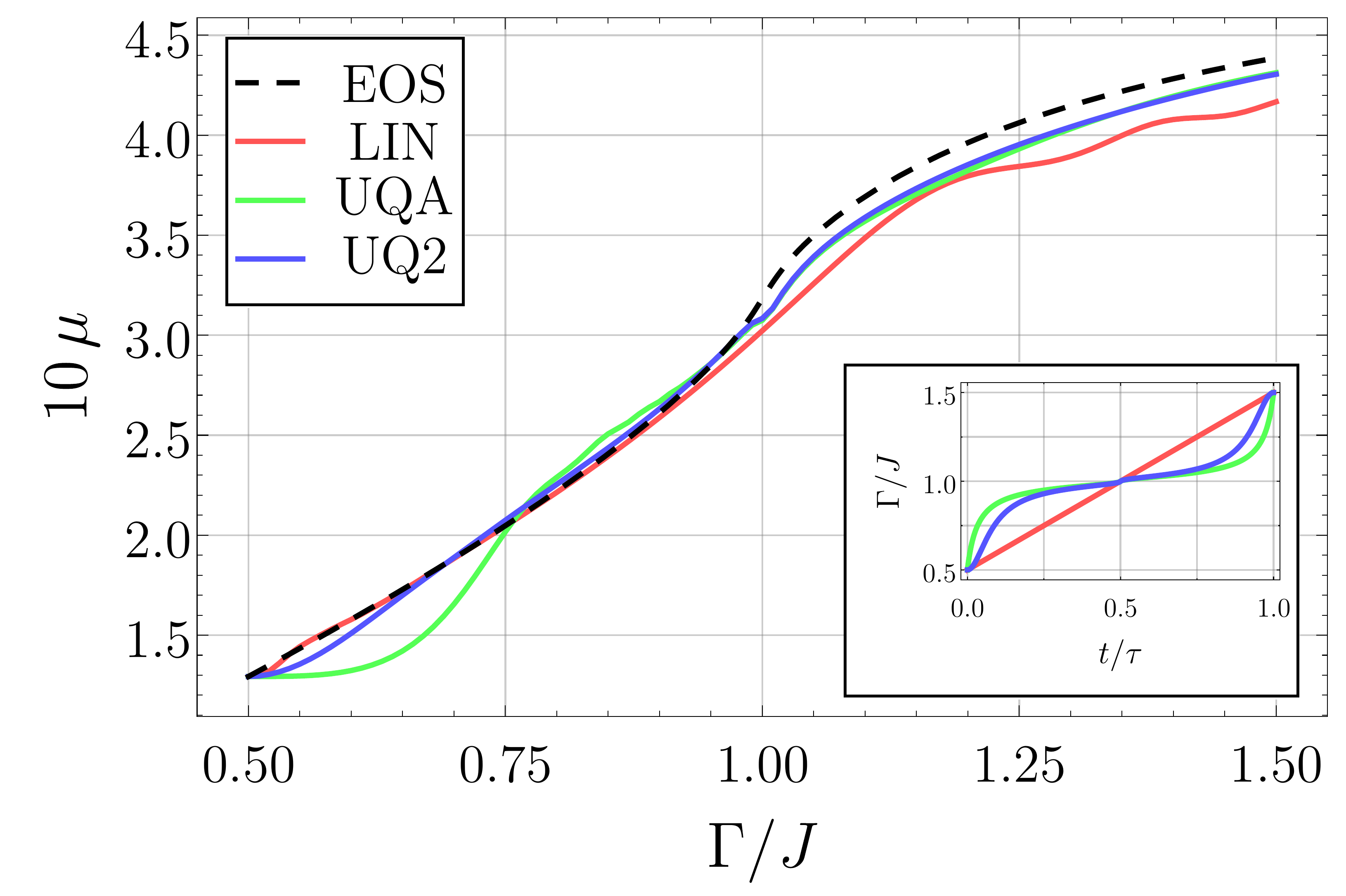}}

\caption{\label{fig:TI_backward}
State diagrams of the time-reversed versions of the processes considered in the main text for the TI chain initially prepared in its ground state with $N = 100$.
The insets show the time-dependence of each protocol.
(a) Reverse paramagnetic process for $J \tau = 3$.
(b) Reverse QCP crossing process for $J \tau = 50$.
}

\end{figure*}

In any process, the linear protocol (LIN) is
\begin{equation} \label{eq:TI_LIN}
\Gamma_{\mathrm{LIN}}(t) = \Gamma_i + \left( \Gamma_f - \Gamma_i \right) \frac{t-t_i}{\tau}.
\end{equation}
The FQA protocol \cite{Martinez2015} is obtained by solving Eq.~(6) of the main text for the two lowest energy states,
\begin{equation} \label{eq:TI_FQA}
\Gamma_{\mathrm{FQA}}(t) = J \cos k_0 + J \sin k_0 \frac{\alpha(t)}{\sqrt{1 - \alpha^2(t)}},
\end{equation}
where $k_0 = \pi/N$,
\begin{equation} \label{eq:TI_FQAalpha}
\alpha(t) = \cos \theta_{k_0,i} + \left( \cos \theta_{k_0,f} - \cos \theta_{k_0,i} \right) \frac{t-t_i}{\tau}
\end{equation}
and
\begin{equation} \label{eq:TItheta}
\theta_k(\Gamma) = \arctan\left( \frac{J \sin k}{\Gamma - J \cos k} \right).
\end{equation}
Equation~(7) of the main text cannot be solved analytically in this case, so we solved it numerically to obtain FQ2.

The UQA \cite{Quan2010} and UQ2 protocols are obtained, respectively, by solving Eqs.~(6) and (7) of the main text with the substitution $F_{mn}(\lambda) \to \pd_{\lambda} E_{mn}(\lambda)$.
For $\Gamma_f < \Gamma_i$, their explicit forms are

\begin{widetext}

\begin{equation} \label{eq:TI_UQA}
\Gamma_{\mathrm{UQA}}(t) = J \left\{ \begin{array}{lc}
\cos k_0 + \sin k_0 \sqrt{ \left( \sin\theta_{k_0,i} + \left( 1 - \sin\theta_{k_0,i} \right) \dfrac{t - t_i}{ t_1 - t_i } \right)^{-1} - 1 }, & t < t_1; \\
\cos k_0 - \sin k_0 \sqrt{ \left( \sin\theta_{k_0,f} + \left( 1 - \sin\theta_{k_0,f} \right) \dfrac{t_f - t}{ t_f - t_1 } \right)^{-1} - 1 }, & t > t_1;
\end{array} \right.
\end{equation}
and
\begin{equation} \label{eq:TI_UQ2}
\Gamma_{\mathrm{UQ2}}(t) = J \left\{ \begin{array}{lc}
\cos k_0 + \sin k_0 \sqrt{ \csc^2 \theta_{k_0,i} \exp\left( \sqrt{2}\, \mathrm{erf}^{-1}\left[ \frac{t - t_i}{ t_2 - t_i } \mathrm{erf}\left( \sqrt{ \log \sin\theta_{k_0,i} } \right) \right] \right)^2 - 1 }, & t < t_2; \\
\cos k_0 - \sin k_0 \sqrt{ \csc^2 \theta_{k_0,f} \exp\left( \sqrt{2}\, \mathrm{erf}^{-1}\left[ \frac{\epsilon_{k_0,f}}{\epsilon_{k_0,i}} \frac{t_f - t}{ t_f - t_2 } \mathrm{erf}\left( \sqrt{ \log \sin\theta_{k_0,i} } \right) \right] \right)^2 - 1 }, & t > t_2; \\
\end{array} \right.
\end{equation}

\end{widetext}

In Eq.~~\eqref{eq:TI_UQ2}, $\mathrm{erf}(z) = \frac{2}{\sqrt{\pi}} \int_0^z e^{-x^2} dx$ is the error function and $\mathrm{erf}^{-1}$ is its inverse.
The times when UQA and UQ2 cross the QCP are $t_1$ and $t_2$ respectively, and they are given by weighted arithmetic mean values of $t_i$ and $t_f$.
For UQA, the weights are $1/(1 - \sin\theta_{k_0})$, while for UQ2, they are $\epsilon_{k_0}/\mathrm{erf}\left( \sqrt{ \log \sin\theta_{k_0} } \right)$.

Figure \ref{fig:TI_backward} shows the state diagrams of the reverse processes starting in the ground state.
In the entirely paramagnetic process of Fig.~\ref{fig:TI_backward_para_EOS}, we can see that FQA outperforms FQ2, because it naturally has a small time derivative at $t_i$.
Accordingly, it has a small first-order APT correction to the magnetization while still doing better than FQ2 at assuring APT, which makes it follow the EOS closely.
This shows that FQA and FQ2 are complementary strategies: if the initial derivative of FQA is small, use it; otherwise, use FQ2.
Conversely, Fig.~\ref{fig:TI_backward_crit_EOS} shows that FQA-like strategies will always have large $\dot{\lambda}_i$ when crossing a critical point, which makes FQ2-like strategies desirable for closely following the EOS.

\paragraph*{Lewis-Riesenfeld invariants: the parametric harmonic oscillator}

\begin{figure*}

\subfloat[\label{fig:HO_forward_EOS}]{\includegraphics[width=.5\textwidth]{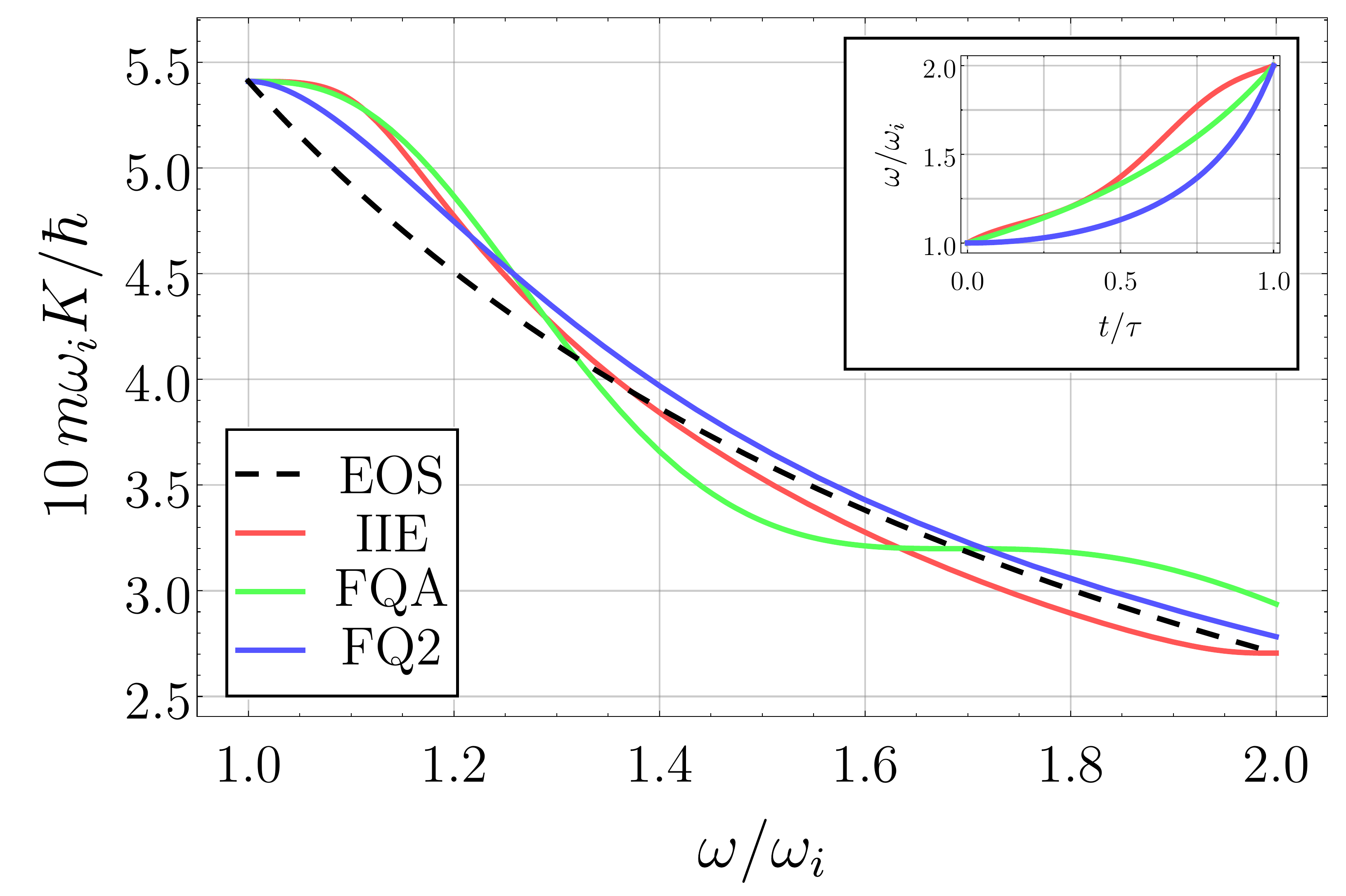}}
\subfloat[\label{fig:HO_backward_EOS}]{\includegraphics[width=.5\textwidth]{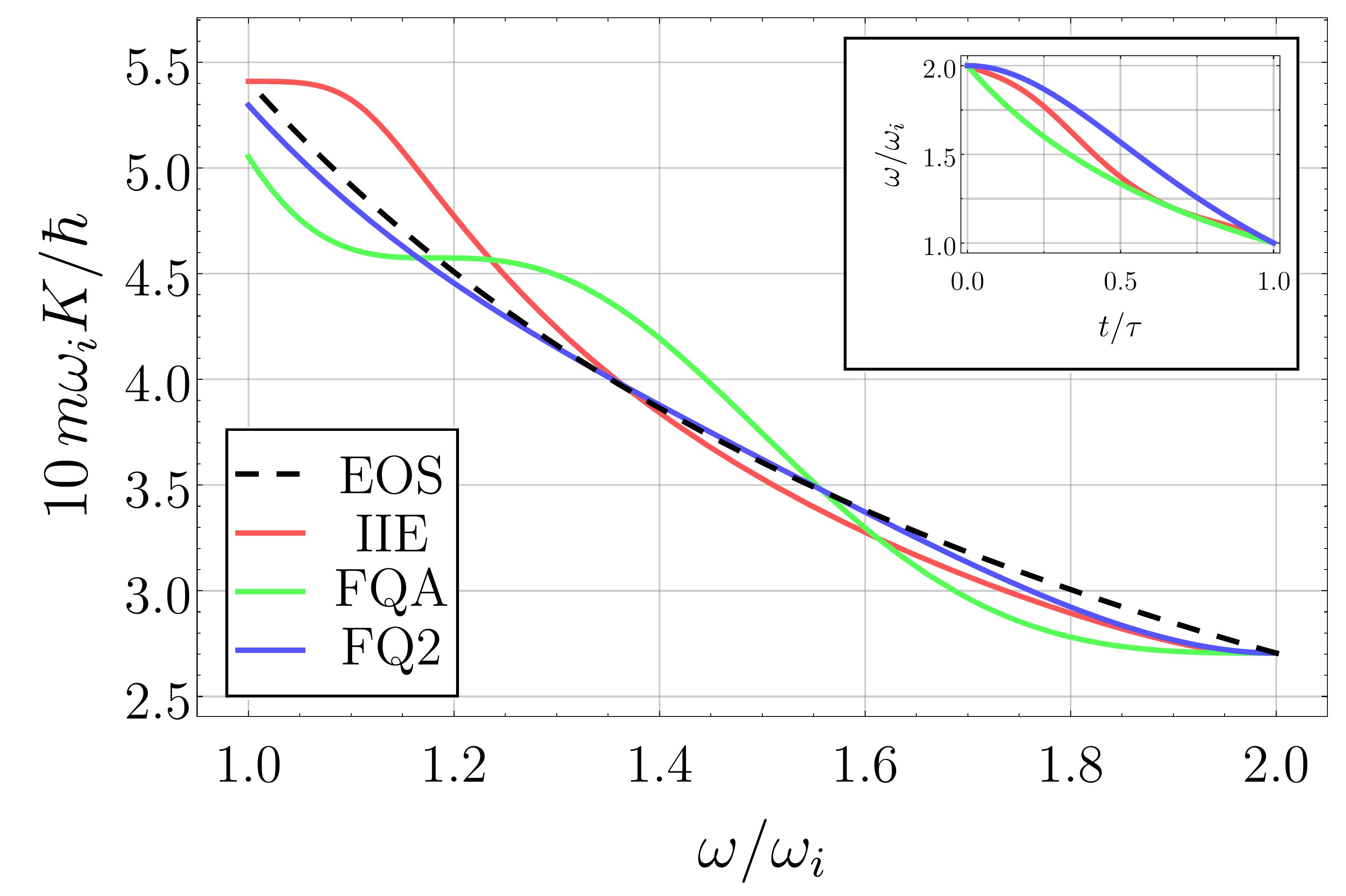}}

\caption{\label{fig:HO_EOS}
State diagrams of the HO for an adiabatic (quasistatic) evolution (EOS), the IIE, the FQA and the FQ2 protocols for $\omega_i \tau = 3$ and $\beta_i \hbar \omega_i = 1$.
The inset shows the form of the protocols.
(a) forward process, starting in the top left corner;
(b) backward process, starting in the the bottom right corner.
}

\end{figure*}

The harmonic oscillator (HO) Hamiltonian, with mass $m$ and varying frequency $\omega$, is
\begin{equation}
 \label{eq:HOHamiltonian}
H_{\mathrm{HO}}(\omega) = \frac{p^2}{2m} + \frac{k(\omega) q^2}{2}, \qquad k(\omega) = m \omega^2.
\end{equation}
The generalized force reads $F_{\mathrm{HO}} = \pd_{k} H_{\mathrm{HO}} = q^2/2$, defined without the minus sign for convenience.
The (non-equilibrium) state variable $K$, conjugate to $k$, can be calculated exactly from
\begin{equation}
 \label{eq:HOStateVariable}
K(t) = \frac{Y^2(t) + \omega_i^2 X^2(t)}{2m \omega_i \omega(t)} \avg{H_{\mathrm{HO}}(\omega(t))}^{(0)},
\end{equation}
where $\avg{H_{\mathrm{HO}}(\omega(t))}^{(0)}$ is the average energy calculated in the adiabatic limit, while $X$ and $Y$ are solutions to the classical equation of motion, $\ddot{Z} + \omega^2 Z = 0$, with initial conditions $X_i = 0 = \dot{Y}_i$ and $\dot{X}_i = 1 = Y_i$ \cite{Husimi1953,Deffner2008,Deffner2010}.

The energetic cost of a given process, quantified as extra work above the quasistatic work, is
\begin{equation}
 \label{eq:HO_work}
W_\mathrm{ex}(t) = \left( Q^*(t) - 1 \right) \avg{H_{\mathrm{HO}}(\omega(t))}^{(0)},
\end{equation}
where (omitting the time-dependences of $X$, $Y$ and $\omega$)
\begin{equation}
\label{eq:HO_Qstar}
Q^*(t) = \frac{\dot{Y}^2 + \omega^2 Y^2 + \omega_i^2 \left( \dot{X}^2 + \omega^2 X^2 \right) }{2\omega_i \omega}
\end{equation}
is an adiabatic measure of the harmonic oscillator: whenever $Q^*(t) = 1$, the system is in the same state as an adiabatic process at time $t$.

The invariant-based inverse engineering (IIE) approach exploits the fact that the system's Hamiltonian admits a Lewis-Riesenfield invariant \cite{Lewis1969} of the form
\begin{equation}
I(t) = \frac{\left( b(t)p - m \dot{b}(t) q \right)^2}{2m} + \frac{m \omega_0^2 q^2}{2b^2(t)},
\end{equation}
as long as $b(t)$ solves the Ermakov equation
\begin{equation} 
\label{eq:HOInvariantEqDif}
\ddot{b}(t) + \omega^2(t) b(t) = \frac{\omega_0^2}{b^3(t)}
\end{equation}
and $\omega_0$ is an arbitrary constant.

Thus, the eigenstates of $I(t)$ are solutions to Schrödinger's equation. One then sets $\omega_0 = \omega_i$ and chooses $b(t)$ such that the eigenstates of $I(t)$ and of $H_{\mathrm{HO}}(\omega)$ coincide at $t_i$ and $t_f$ (up to irrelevant phases) \cite{Chen2010}. The corresponding protocol $\omega_{\mathrm{IIE}}(t)$ is obtained from Eq.~\eqref{eq:HOInvariantEqDif}, and it guarantees, for any process duration $\tau$, the same final state as an adiabatic process.
There is freedom in choosing the exact form for $b(t)$ --- we chose the simplest polynomial in $t/\tau$ that satisfies the required boundary conditions.
Note that, while $b(t)$ is a strict function of $t/\tau$, $\omega_{\mathrm{IIE}}(t)$ obtained from Eq.~\eqref{eq:HOInvariantEqDif} is not.
We shall see later on that the IIE protocol has some unique properties on the state diagram, owning to its invariant-based design.

Solving Eq.~(6) of the main text for the HO with the proper boundary conditions gives us the FQA protocol \cite{Martinez2015}
\begin{equation} 
\label{eq:HO_FQAprotocol}
\omega_{\mathrm{FQA}}(t) = \left( \frac{1}{\omega_i} + \left( \frac{1}{\omega_f} - \frac{1}{\omega_i} \right) \frac{t-t_i}{\tau} \right)^{-1}.
\end{equation}
On the other hand, solving Eq.~(7) of the main text with $\dot{\omega}_i = 0$ gives us the FQ2 protocol
\begin{equation} 
\label{eq:HO_FQ2protocol}
\omega_{\mathrm{FQ2}}(t) = \omega_i \exp\left\{ \mathrm{erf}^{-1}\left[ \frac{t-t_i}{\tau} \mathrm{erf}\left( \sqrt{ \log\frac{\omega_f}{\omega_i} } \right) \right] \right\}^2.
\end{equation}

Figure \ref{fig:HO_EOS} shows forward and backward processes for the same $\tau$ and an initial canonical distribution of inverse temperature $\beta_i$.
Note in the insets that the forms of IIE and FQA in the backward process are simply their time-reversed forms of the forward process, but this is not the case for FQ2, given its requirement to set $\dot{\omega}_i = 0$ while not setting $\dot{\omega}_f$.
We can clearly see that, out of the three protocols, FQ2 does the best job of keeping the evolution closer to the EOS at all times.
Of course, FQA and FQ2 do not have the same exact final state predicted by the EOS, unlike IIE, which requires more information about the system and is, therefore, inaccessible in most cases.
Peculiarly, IIE traces the exact same curve of the state diagram in the forward process (FIG.~\ref{fig:HO_forward_EOS}) and in the backward process (FIG.~\ref{fig:HO_backward_EOS}), which is a consequence of its invariant-based nature.

\begin{figure*}

\subfloat[\label{fig:HO_forward_force}]{\includegraphics[width=.5\textwidth]{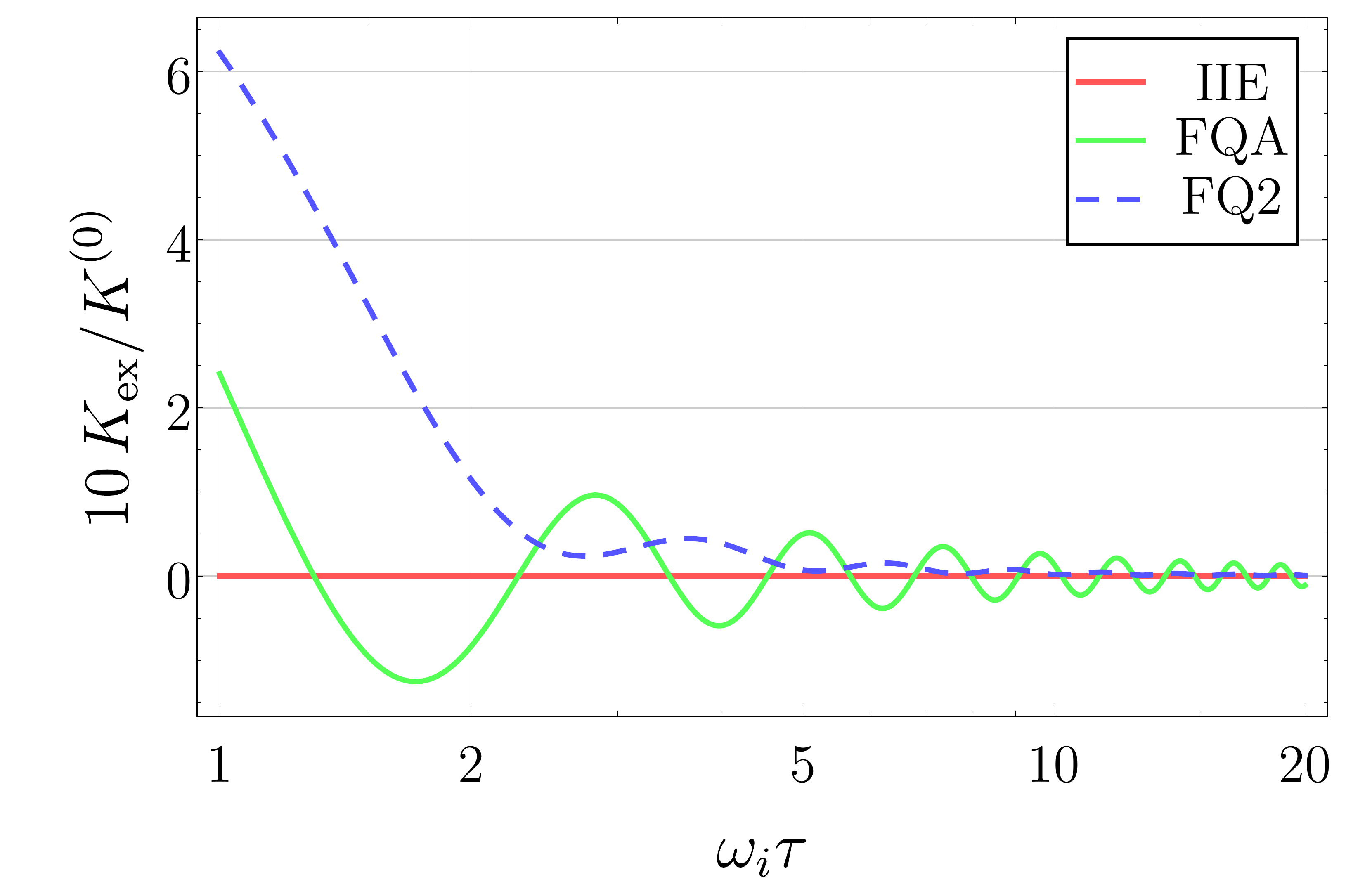}}
\subfloat[\label{fig:HO_forward_work}]{\includegraphics[width=.5\textwidth]{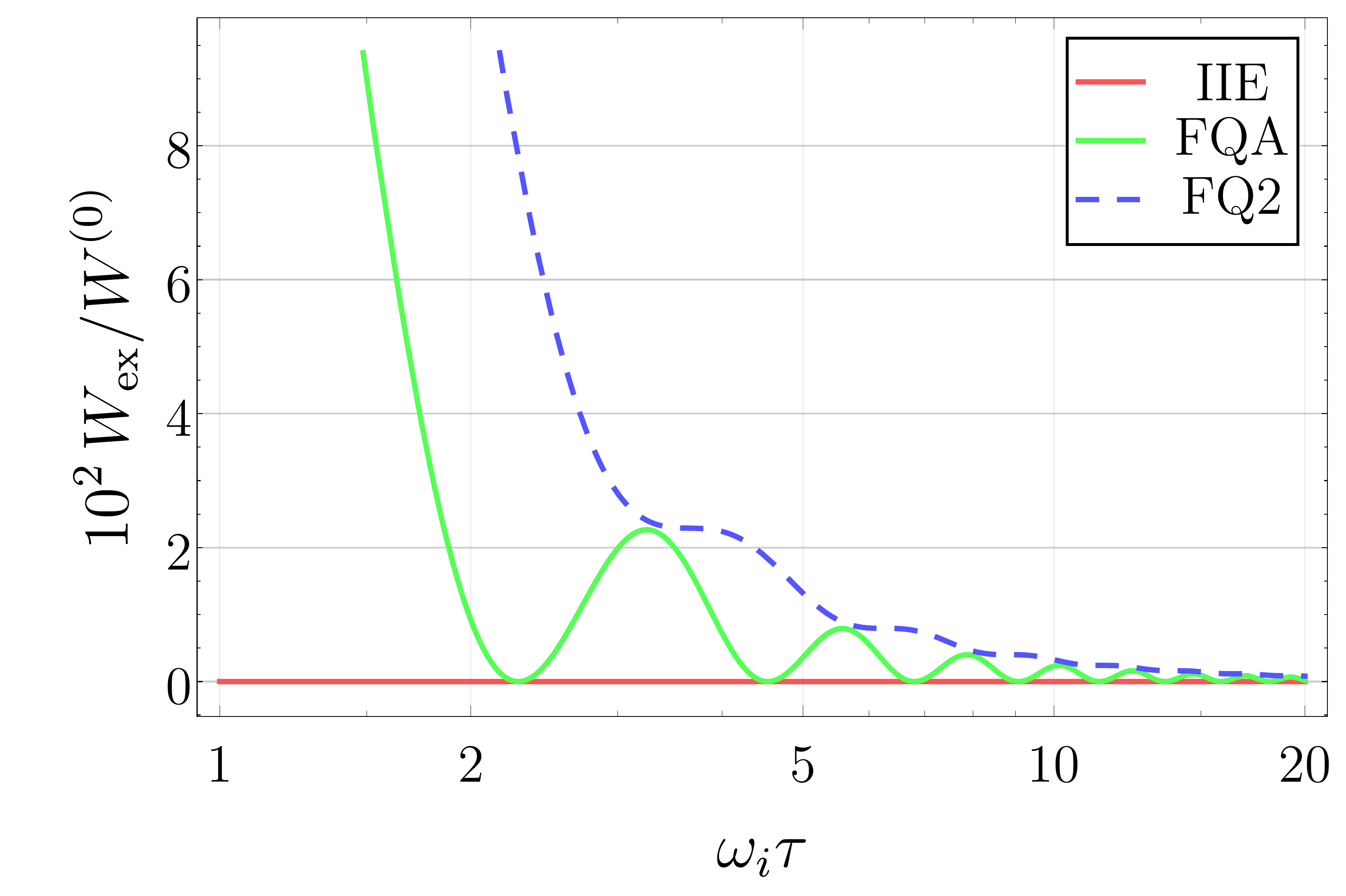}}

\caption{\label{fig:HO_forward_forceANDwork}
(a) Excess state variable vs. $\tau$ for the HO forward process.
(b) Excess work vs. $\tau$ for the HO forward process.
}

\end{figure*}

Figure~\ref{fig:HO_forward_forceANDwork} shows, at $t_f$, the excess state variable and excess work vs. $\tau$ in the forward process.
We see, from Fig.~\ref{fig:HO_forward_force}, that FQA achieves $K_{\mathrm{ex}} = 0$ for a specific small time in the range $1 < \omega_i \tau < 2$, while FQ2 gives an overall smaller $K_{\mathrm{ex}}$ for an arbitrary $\omega_i \tau > 2$.
This is consistent with their definitions: the FQA protocol of Eq.~\eqref{eq:HO_FQAprotocol} does a better job at securing an accurate description of the microscopic dynamics by means of APT, but it does not necessarily give the smallest deviations from the adiabatic theorem.
On the other hand, the FQ2 protocol of Eq.~\eqref{eq:HO_FQ2protocol} gives up some (but not much) of its ability to attain early (for small $\tau$) adiabaticity in order to allow better following of the EOS.
Of course, the IIE protocol is built to give the same final state as an adiabatic evolution, and thus gives $K_{\mathrm{ex}} = 0$ for any $\tau$.

\begin{figure}

\includegraphics[width=\columnwidth]{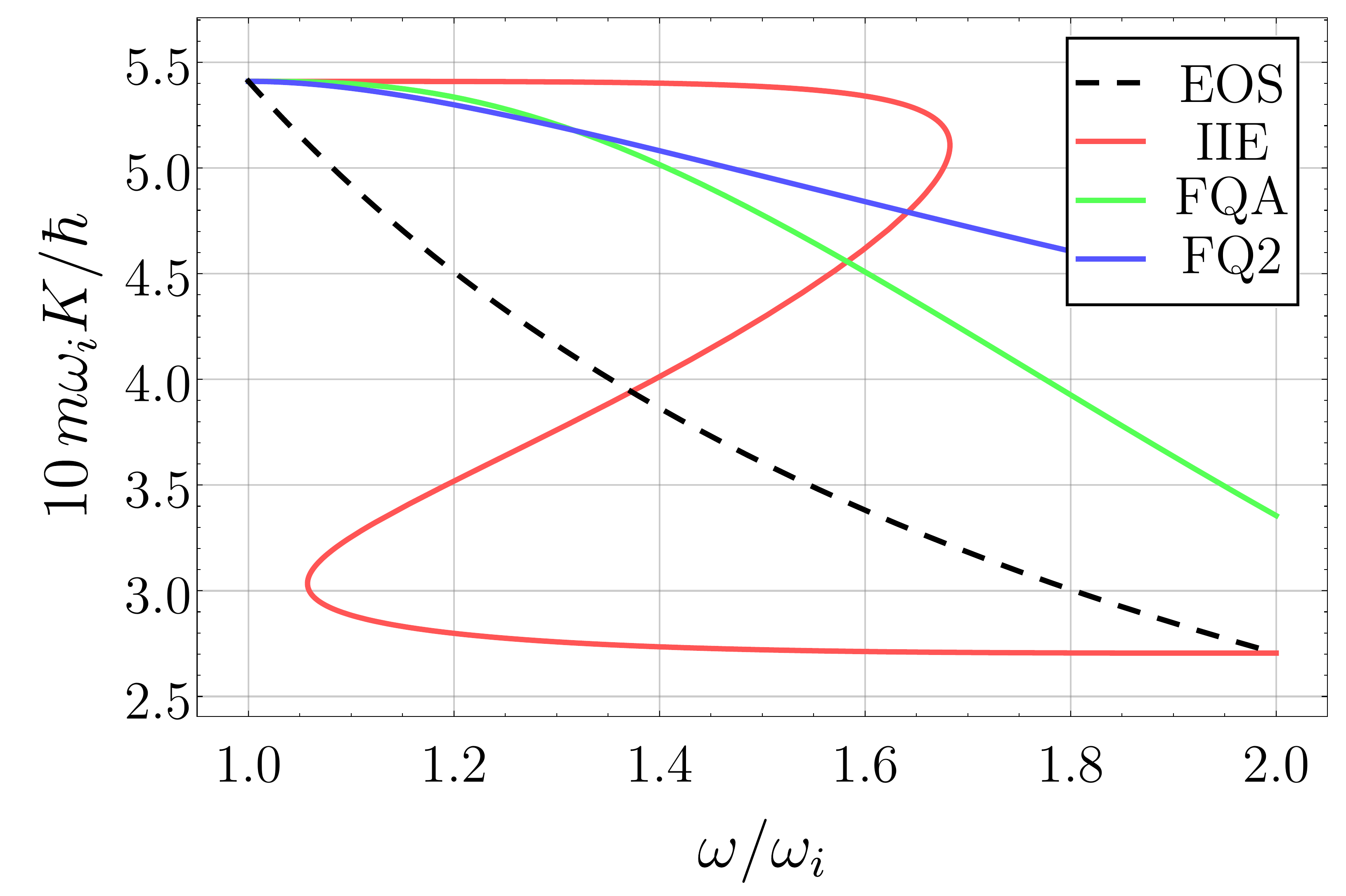}

\caption{\label{fig:HO_forward_EOS1}
State diagram for the HO forward process with $\omega_i \tau = 1$, starting from the top left corner.
}

\end{figure}

Figure~\ref{fig:HO_forward_work} clearly shows that, when it comes to the energetic cost of the forward process, FQA beats FQ2 for any $\tau$.
Thus, in order to closely follow the EOS, one might have to spend more energy throughout the process.
Interestingly, IIE gives also $W_{\mathrm{ex}} = 0$ for any $\tau$ and, in fact, this can be inferred from the state diagram given in Fig.~\ref{fig:HO_forward_EOS}.
The total excess work of a process for a given protocol is the area between its state variable curve and the EOS curve in a state diagram, and we can see that IIE crosses the EOS exactly one time (at $t = (t_f + t_i)/2$).
Thus, any excess energy given to the system in the first half of the process (area above the EOS) is retrieved in the second half (area below the EOS), netting zero excess work.
This must be the case for any $\tau$, even if the evolution is far from being adiabatic at all times.
To illustrate this, we have included Fig.~\ref{fig:HO_forward_EOS1}, where the effects just described are naturally exacerbated.
All in all, IIE accomplishes ``finite-time reversibility'' when the whole process is taken into account, re-treading its thermodynamic path when time is reversed and dissipating zero energy.

\bibliography{bibliography}